\documentclass[%
 twocolumn,
 amsmath, amssymb,
 aps,
 pre,
 longbibliography
]{revtex4-1}

\usepackage{graphicx}
\usepackage{dcolumn}
\usepackage{bm}
\usepackage[english]{babel}
\usepackage[pdfpagelabels,plainpages=false,bookmarks=true,colorlinks,linkcolor=red,urlcolor=blue,citecolor=blue]{hyperref}

\usepackage{multirow}
\usepackage{diagbox}
\renewcommand{\vec}[1]{{\bf #1}}
\renewcommand{\Re}[1]{{\text{Re}\{#1\}}}

\newcommand{\braket}[1]{\langle #1  \rangle}
\newcommand{\ket}[1]{| #1  \rangle}

\newcommand{\norm}[1]{\left\lVert#1\right\rVert}

\newcommand{\Potts}{\text{Potts}}

\newcommand{\Log}{\text{Log}}
\newcommand{\abs}[1]{|#1|}

\newcommand{\rg}[1]{{(#1)}}
\newcommand{\tmp}{\text{tmp}}
\newcommand{\Tr}{\text{Tr}}

\newcommand{\T}{\vec{T}}
\newcommand{\X}{\vec{X}}
\renewcommand{\u}{\vec{u}}
\renewcommand{\v}{\vec{v}}

\newcommand{\A}{\vec{A}}

\begin{document}

\title{Dynamical Quantum Phase Transitions of Quantum Spin Chains with the Loschmidt-rate Critical Exponent equal to $\frac{1}{2}$}

\author{Yantao Wu}

\affiliation{
The Department of Physics, Princeton University
}
\date{\today}
\begin{abstract}
We describe a new universality class of dynamical quantum phase transitions of the Loschmidt amplitude of quantum spin chains off equilibrium criticality. 
We demonstrate that in many cases it is possible to change the conventional linear singularity of the Loschmidt rate function into a smooth peak by tuning one parameter of the quench protocol. 
Exactly at the point when this change-over occurs, the singularity of the Loschmidt rate function persists, with a critical exponent equal to $\frac{1}{2}$. 
The non-equilibrium renormalization group fixed-point controlling this universality class is described. 
An asymptotically exact renormalization group recursion relation is derived around this fixed-point to obtain the critical exponent.  
\end{abstract}

\pacs{Valid PACS appear here}
\maketitle
\section{Introduction}
In recent years, there has been a surge of interest in the post-quench dynamics of a quantum system, due to the rapid development in experimental techniques \cite{Kinoshita2006,Gring1318,Blatt2012,RevModPhys.86.153,Bloch2012,RevModPhys.80.885} and numerical algorithms \cite{tebd,itebd,tdvp,PhysRevB.94.165116,expH_MPO}.  
In particular, critical phenomena are found to appear in the post-quench out-of-equilibrium dynamics of quantum systems.    
In the seminal paper \cite{DQPT}, a notion of dynamical quantum phase transition (DQPT) is identified in the Loschmidt amplitude $G(t)$ of a quantum quench of the transverse field Ising model:   
\begin{equation}
  G(t) = \braket{\psi_0 | e^{-i\hat H t}|\psi_0} = \braket{\psi_0 |\psi(t)}
\end{equation}
which measures the return probability of a quantum state $\ket{\psi_0}$ under the time evolution of Hamiltonian $\hat H$. 
In general, $G(t)$ satisfies a large-deviation principle \cite{DQPT,Large_Deviation} and its rate function $l(t)$ is intensive in the thermodynamic limit, 
\begin{equation}
  l(t) = -\lim_{L\rightarrow \infty} \frac{1}{L} \ln \abs{G(t)}^2 = -\lim_{L\rightarrow \infty} \frac{2}{L} \Re{\Log{G(t)}}
\end{equation}
where $L$ is the system size, and $\Log$ is the principal complex logarithmic function. 
\cite{DQPT} found $l(t)$ to be singular at certain critical times.  
Later on, people have discovered many examples of DQPTs, e.g. \cite{Heyl_2018, Zvyagin_2016, DQPT_disentangle, DQPT_XY, DQPT_XXZ, DQPT_longrange,DQPT_Potts, DQPT_Fisherzero, DQPT_concurrence,DQPT_nonintegrable,DQPT_transfer,DQPT_nnn_Ising,DQPT_firstorder,DQPT_nonintegrable_dutta,DQPT_merging,DQPT_slowquench,DQPT_semiclassical,DQPT_extendedising,DQPT_domainwall}, investigating different aspects of them. 
Significant progress has also been made on their experimental observation \cite{experiment1,experiment2}.  

A major goal of quantum dynamical critical phenomena is to classify all possible non-analyticities of the Loschmidt rate function.   
In the most common cases, one looks for power-law singularities at critical times $t_c$, in the form  
\begin{equation}
  l(t) = A_\pm \abs{t - t_c}^{\alpha_\pm} + \text{reg.} 
\end{equation}
where $\text{reg.}$ is a regular function of $t$. 
Here $A_\pm$ and $\alpha_\pm$ are the critical amplitude and exponent of the singularity, with $+$ standing for $t > t_c$ and $-$ for $t < t_c$. 
As in the equilibrium critical phenomena, $\alpha_\pm$ is universal and robust against small perturbations, and is thus the central quantity in characterizing a universality class of DQPTs.  
In a recent study of a disordered many-body localized chain, a critical exponent of approximately 0.2 was numerically observed \cite{DQPT_MBL}. 
However, to our best knowledge, almost all DQPTs discovered in one dimensional pure quantum spin chains exhibit linear singularities in $l(t)$, i.e., $\alpha_+ = \alpha_- = 1$, which can be understood through the transfer matrix formalism \cite{DQPT_transfer,DQPT_Fisherzero}.    
For a pure quantum spin chain, $G(t)$ can be described by a product of $L$ transfer matrices $\vec T(t)$:  
\begin{equation}
  G(t) = \Tr(\T(t)\T(t)\cdots),   
  \label{eq:G_transfer}
\end{equation}
where $\vec T(t)$ depends smoothly on $t$.  
Let the dimension of $\vec T(t)$ be $D$, which can be finite or infinite. 
In the thermodynamic limit, the Loschmidt rate function is then given by the leading eigenvalue of $\vec T(t)$:
\begin{equation}
  \begin{split}
    l(t) &= \lim_{L \rightarrow \infty} -\frac{2}{L} \Re{\Log \sum_{i=1}^D \lambda_i} 
\\ 
&=\lim_{L\rightarrow \infty} -\frac{2}{L} \Re{\Log\lambda_{\max}^L} - \frac{2}{L} \Re{ \Log\sum_{i=1}^D (\frac{\lambda_i}{\lambda_{\max}})^L}
 \\
 &=-2 \max_{i=1, \cdots, D} \ln \abs{\lambda_i(t)}.
\end{split}
\end{equation}
where $\lambda_i$s are the eigenvalues of $\T$, and $\lambda_{\max}$ is the eigenvalue with the largest modulus. 
If the two largest eigenvalues of $\vec T(t)$ are both non-degenerate, then they depend smoothly on the matrix elements of $\vec T(t)$, and thus on $t$.
In this case, the singularity in $l(t)$ occurs when the modulus of these two eigenvalues equal, and generically exhibits a critical exponent equal to one.  
For quantum spin chains, an important unanswered question is thus whether there are universality classes of DQPTs with $\alpha_\pm$ different from one \cite{Heyl_2018}.  

In this paper, we demonstrate that DQPTs in quantum spin chains with critical exponent $\frac{1}{2}$ also occur generically, through analysis of the renormalization of the transfer matrix.    
The transfer matrix can be obtained, for example, in matrix product states (MPS)-based time evolution algorithms as \cite{itebd, tdvp, DQPT_transfer, DQPT_Fisherzero}
\begin{equation}
  \T(t) = \sum_{s} \overline{\A}^s_0\otimes \A^s(t) 
  \label{eq:MPS}, 
\end{equation}
with $s$ indexing the physical degree of freedom at a local lattice site. 
Here, $\A^s_0$ and $\A^s(t)$ are respectively the matrix in the MPS representation of $\ket{\psi_0}$ and $\ket{\psi(t)}$.  
Off quantum criticality, although the entanglement entropy of a quantum spin chain generally increases with time \cite{Entanglement_dynamics}, it still is finite at a finite $t$, making a finite dimensional $\vec T(t)$ possible \cite{mera,tebd}, as evidenced by the many numerical results obtained from MPS-based time evolutions algorithms \cite{DQPT_longrange, DQPT_Potts, DQPT_Fisherzero, DQPT_concurrence, DQPT_nonintegrable, DQPT_nnn_Ising, DQPT_domainwall}.    
When $\ket{\psi_0}$ is at equilibrium criticality, however, a finite-dimensional $\vec T(t)$ will not be available and our analysis below will be invalid. 
In the following we will use ``$\frac{1}{2}$-DQPT'' to denote the DQPTs with exponent $\frac{1}{2}$, and ``linear-DQPT'' to denote those with exponent 1.

While in the literature, to the best of our knowledge, there has been no mentions of $\frac{1}{2}$-DQPTs, it has been observed several times that linear-DQPTs can be made to disappear by changing parameters of the quench protocol \cite{DQPT, DQPT_disentangle, DQPT_XY, DQPT_longrange,DQPT_Potts, DQPT_Fisherzero, DQPT_concurrence}.      
These observations suggest that there exist critical parameters of the quench protocol at which the linear-DQPTs terminate. 
Although not all these termination points give rise to a $\frac{1}{2}$-DQPT, as we will show, if a $\frac{1}{2}$-DQPT does occur, it occurs at the termination point of a linear-DQPT.  

The paper is organized as follows. 
In Sec. \ref{sec:RG}, we present the renormalization group (RG) calculation. 
In Sec. \ref{sec:Potts}, we give a concrete example of the $\frac{1}{2}$-DQPT in the three-state Potts chain.
In Sec. \ref{sec:Discussion}, we discuss and conclude.  

\section{Renormalization Group Calculation}
\label{sec:RG}
\subsection{The RG procedure}
RG has proved a powerful tool to analyze equilibrium phase transitions \cite{wilson_rg}.       
Its utility in DQPTs was first demonstrated by Heyl in \cite{DQPT_RG} which re-explained the linear DQPT in the transverse-field Ising chain through coarse-graining the system Hamiltonian by the decimation rule.       
We recently generalized Heyl's RG procedure to the coarse-graining of the transfer matrix of $G(t)$, which avoided a lot of the mathematical complication of the complex logarithmic function \cite{Potts_RG}.   
We review the RG procedure here.  
To analyze $l(t)$, we perform the decimation coarse-graining \cite{cardy}, i.e. every other spin is summed away.   
The decimation coarse-graining is equivalent to multiplying two neighboring transfer matrices into one.  
The renormalized transfer matrix $\T^{(n+1)}$ at the $(n+1)$th RG iteration is thus given by 
\begin{equation}
  \begin{split}
  \text{step 1: }& \T^{(n+1)}_\tmp = \T^{\rg{n}}\T^{\rg{n}} 
  \\
  \text{step 2: }& \T^{\rg{n+1}} = \frac{\T^\rg{n+1}_\text{tmp}}{(\T^\rg{n+1}_\text{tmp})_{11}}
\end{split}
\label{eq:RG}
\end{equation}
where step 2 isolates out the overall multiplicative growth of $\T^\rg{n}$ and is necessary for the RG fixed-points to exist.   
For notational consistency, we define $\T^\rg{0}$ as the unrenormalized transfer matrix. 
For a finite chain of length $L$, $\log_2 L$ number of RG iterations can be carried out, and the product of $\abs{(\T^\rg{n}_\tmp)_{11}}$ extracted at each RG step gives the value of the Loschmidt amplitude.   
Thus, the RG procedure also provides an expression of the Loschmidt rate function:  
\begin{equation}
  \begin{split}
    l(t) &= -\lim_{L \rightarrow \infty} \frac{2}{L} \sum_{n=1}^{\log_2L} \ln\abs{(\T_\tmp^\rg{n}(t))_{11}}^{\frac{L}{2^n}} 
  \\
  &= -\sum_{n=1}^\infty \frac{1}{2^{n-1}} \ln\abs{(\T_\tmp^\rg{n}(t))_{11}}. 
\end{split}
  \label{eq:Tnn}
\end{equation}

Here we pause to comment on the normalization choice in Eq. \ref{eq:RG}, which may appear quite arbitrary. 
Even if $\ket{\psi_0}$ and $\ket{\psi(t)}$ are normalized to unit norm, the leading eigenvalue of the transfer matrix will in general have a non-unit modulus, which gives the very phenomena of DQPT. 
Thus, if no normalization is done, no RG fixed-point will exist. 
When one extracts the normalization factors to compute the rate function, as in Eq. \ref{eq:Tnn}, one also needs to demand the factor extracted at each RG iteration be an analytic function of $t$.  
This is very important in using the RG analysis to isolate out the singular behavior of $l(t)$ in order to compute the critical exponent, and is also enforced in RG calculations of equilibrium phase transitions. 
This criterion rules out using eigenvalues of $\T_\tmp$ to normalize, as eigenvalues are in general not analytic functions of $t$.  
The matrix elements of $\T^\rg{n}_\tmp$, however, depend analytically on $t$, because they are just finite combinations of addition, multiplication, and division of the matrix elements of the unrenormalized transfer matrix, which all depend analytically on $t$. 
There still remains the arbitrariness in which matrix element one should use to normalize the RG procedure. 
This is indeed arbitrary, and any matrix element should work, as long as they do not become zero in the RG flow. 
Here, we assume that the matrix element $(\T^\rg{n+1}_{\text{tmp}})_{11}$ never becomes zero. 
If it does, in Eq. \ref{eq:RG} and Eq. \ref{eq:Tnn}, we can replace $(\T^\rg{n+1}_\tmp)_{11}$ with a non-zero $(\T^\rg{n+1}_{\text{tmp}})_{12}$, and so on.  
However, when $\vec T^\rg{n+1}_\tmp$ becomes a zero matrix, no replacement can be done and the sum in Eq. \ref{eq:Tnn} will develop a singularity. 
As we will see, this is exactly what happens at the RG fixed-point of a $\frac{1}{2}$-DQPT.   

\subsection{RG fixed-point} 
Consider transfer matrices $\T(t)$ of dimension $D \times D$ which depend smoothly on time. 
Let a transfer matrix be written as   
\begin{equation}
  \T = \begin{pmatrix} 1 & \vec v^{T} \\ \vec u & \vec X \end{pmatrix}
    \label{eq:T}
\end{equation}
where $\vec u$ and $\vec v$ are column vectors of dimension $D-1$ and $\X$ is a $(D-1)\times(D-1)$ matrix. 
Because the RG procedure is analogous to a power iteration of $\T$, a fixed-point transfer matrix exists only if there is a unique leading eigenvector.     
In that case, the fixed-point transfer matrix $\T^*$ will be 
\begin{equation}
  \T^* = \begin{pmatrix} 1 & \vec v^{*T} \\ \vec u^* & \vec X^* \end{pmatrix}
\end{equation}
where $(1, \vec u^{*T})^T$ is the leading right eigenvector of $\vec T$ and $(1, \vec v^{*T})$ is the leading left eigenvector. 
In addition, 
\begin{equation}
  \u^* \v^{*T} = \X^*.   
  \label{eq:condition}
\end{equation}
Note that $\T^*$ satisfies the fixed-point equation of Eq. \ref{eq:RG}:  
\begin{equation}
  \T^* \T^* = (1 + \v^{*T} \u^*) \T^*. 
  \label{eq:fp_equation}
\end{equation}
In a linear-DQPT, there is a discontinuous jump of the leading eigenvector at $t_c$ and thus the matrix element of $\T^*(t)$ will also experience a discontinuous jump as a function of $t$.  
Indeed this behavior is seen, for example, for the Potts chain studied later (Fig. \ref{fig:rg_itebd}, left panel).  
In Eq. \ref{eq:T}, we normalized the transfer matrix so that $\vec T_{11} = 1$. 
In retrospect, this normalization is valid in the entire RG flow if the first elements of both the leading right-eigenvector and the leading left-eigenvector of $\vec T$ are non-zero. 
If not, we can normalize $\vec T_{12}$ to be 1, and so on, and change the normalization of the RG procedure accordingly.  
\subsection{RG flow at the $\frac{1}{2}$-DQPT}
When the parameters of the quench protocol are varied so that the smooth time dependence of $\T^*(t)$ changes into a discontinuity, this could suggest  a singularity in the RG procedure, i.e., $\T^*_\tmp = \T^* \T^*$ becomes the zero matrix. 
If this occurs, as we now show, the RG fixed-point will give a $\frac{1}{2}$-DQPT. 
In this case, in addition to Eq. \ref{eq:condition}, the $\u_c$ and $\v_c$ of $\T_c$ must be such that (hereinafter, we write $\T_c$ as the fixed-point transfer matrix controlling the $\frac{1}{2}$-DQPT)   
\begin{equation}
1 + \v_c^T\u_c = 0.      
\label{eq:condition2}
\end{equation}
That is, the right-eigenvector and the left-eigenvector of $\vec T$ becomes orthogonal (in the sense of the inner product of the real Euclidean space), suggesting that the leading eigenvalue of the transfer matrix is defective, i.e., the algebraic multiplicity of the leading eigenvalue is larger than its geometric multiplicity \cite{defective}.  

Now assume that at a certain quench parameter $J_c$ and a critical time $t_c$, the transfer matrix approaches a fixed-point transfer matrix $\T_c$ that satisfies Eq. \ref{eq:condition2}. 
When $t$ is close to $t_c$, a finite number of RG iterations will take the transfer matrix to the vicinity of $\T_c$. 
Thus, to determine the universal behavior of the $\frac{1}{2}$-DQPT, one needs to study the RG flow around $\T_c$.    
In a conventional RG analysis, one assumes that the RG equation is analytic at the critical fixed-point and linearizes the RG flow around it \cite{cardy}.   
However, in our case, the RG procedure becomes non-analytic precisely at $\T_c$. 
As a result, the naive expansion $\u = \u_c + \delta \u, \v = \v_c + \delta \v$, and $\X = \X_c + \delta \X$ fails at producing a recursion relation of $\delta \u, \delta \v$, and $\delta \X$ to the leading order.      
The trick, inspired by Eq. \ref{eq:condition}, is to do the small-parameter expansion in the following way:   
\begin{equation}
  \T^{(n)} = \begin{pmatrix} 1 & (\v_c + \delta\v^\rg{n})^T \\
    \u_c + \delta \u^\rg{n} & (\u_c + \delta \u^\rg{n})(\v_c + \delta\v^\rg{n})^T + \delta \X^\rg{n}
\end{pmatrix}
\end{equation}
Then assuming Eq. \ref{eq:condition2}, one obtains to the leading order (see the supplementary material (SM) \cite{sm} for the derivation),   
\begin{equation}
  \begin{split}
    \delta \u^\rg{n+1} = \delta \u^\rg{n} + \frac{\delta \X^\rg{n} \, \u_c}{\v_c^T \delta \u^\rg{n} + (\delta \v^\rg{n})^T \u_c} 
\\
(\delta \v^\rg{n+1})^T = (\delta \v^\rg{n})^T + \frac{\v_c^T \, \delta \X^\rg{n}}{\v_c^T \delta \u^\rg{n} + (\delta \v^\rg{n})^T \u_c} 
\\
\delta \X^\rg{n+1} = -\frac{\delta \X^\rg{n} \, \u_c \v_c^T \delta \X^\rg{n}}{(\v_c^T \delta \u^\rg{n} + (\delta \v^\rg{n})^T \u_c)^2}.  
\end{split}
\label{eq:recursion}
\end{equation}
This recursion relation is asymptotically exact in the sense that as $t\rightarrow t_c$ and $n\rightarrow \infty$, $\delta\u,\delta\v,$ and $\delta \X$ become progressively smaller, making the expansion more and more accurate near criticality.  

To connect this recursion relation with the rate function, note that Eq. \ref{eq:Tnn} can be separated into two parts: 
\begin{equation}
  \begin{split}
    l(t) = &-\sum_{n=1}^{n_0} \frac{1}{2^{n-1}} \ln\abs{(\T_\tmp^\rg{n}(t))_{11}} 
  \\
  &-\sum_{n=n_0 + 1}^\infty \frac{1}{2^{n-1}} 
\ln\abs{(\T_\tmp^\rg{n}(t))_{11}}
\end{split}
  \label{eq:parts}
\end{equation}
where $n_0$ is a finite positive integer. 
$n_0$ number of RG iterations would need to be carried out to reach the vicinity of $\T_c$ so that $\norm{\delta \u},\norm{\delta \v}$, and $\norm{\delta \X}$ are all much less than one and Eq. \ref{eq:recursion} can be applicable for $n > n_0$. 
As mentioned, for any finite $n$, $(\T^\rg{n}_\tmp)_{11}$ will be an analytic function of $t$. 
Thus, the first sum in Eq. \ref{eq:parts} is an analytic function of $t$ and will be dropped from the singular part of the rate function, $l_s(t)$. 
Then, to the leading order of $\delta \u$ and $\delta \v$,  
\begin{equation}
  \begin{split}
  l_s(t) &= \sum_{n=n_0}^\infty \frac{-1}{2^n} \ln\abs{\v_c^T\delta\u^\rg{n}(t) + (\delta\v^\rg{n}(t))^T\u_c}.
\end{split}
\label{eq:ls}
\end{equation}
To compute $l_s(t)$, one needs to derive from Eq. \ref{eq:recursion} a recursion relation for $\v_c^T\delta\u^\rg{n}(t) + (\delta\v^\rg{n}(t))^T\u_c$. 
This is technical and is presented in the Appendix (Sec. \ref{sec:proof_ls_w}). 
The result, however, is very simple: there exists a non-universal constant $a$ such that if one defines $\delta{\tilde\omega}^\rg{n} \equiv \frac{\v_c^T\delta\u^\rg{n}(t) + (\delta\v^\rg{n}(t))^T\u_c}{2 \sqrt{a\delta t}}$, 
then
\begin{equation}
l_s(t) =\frac{-1}{2^{n_0-1}}\ln\abs{\sqrt{\delta t}} + \sum_{n=n_0}^\infty \frac{-1}{2^{n}} \ln \abs{\delta \tilde\omega^\rg{n}(t)}
\label{eq:ls_w}
\end{equation}
where $\delta t = t - t_c$.  
The recursion relation of $\delta\tilde{w}^\rg{n}(t)$ is given by  
\begin{equation}
    \delta \tilde w^\rg{n+1} = \frac{1}{2}(\delta \tilde w^\rg{n} + \frac{1}{\delta \tilde w^\rg{n}}), \hspace{2mm} \delta \tilde w^\rg{n_0}(t) = \frac{\delta w_0}{\sqrt{a\delta t}}
    \label{eq:w_recursion}
\end{equation}
where $\delta w_0$ is a non-universal constant. 

It is now possible to look for $t_1$ and $t_2$ such that $\delta \tilde w^\rg{n_0}(t_2) = \delta \tilde w^\rg{n_0 + 1}(t_1)$ so that $\delta \tilde w^\rg{n}(t_2) = \delta \tilde w^\rg{n + 1}(t_1)$ for all $n > n_0$. 
This requires $\frac{\sqrt{\delta t_1}}{\sqrt{\delta t_2}} = \frac{1}{2}(1 + \frac{a\delta t_1}{\delta\omega_0^2})$.
Then the sum in Eq. \ref{eq:ls_w} are related simply for $t_1$ and $t_2$, and one obtains (see the SM \cite{sm} for a derivation) 
\begin{equation}
  \begin{split}
  l_s(t_1) &= \frac{1}{2} l_s(t_2) -\frac{1}{2^{n_0}} \ln\abs{\frac{\sqrt{\delta t_1}}{\sqrt{\delta t_2}}} + C
  \\
  &= \frac{1}{2}l_s(t_2) + O(\delta t_1) 
\end{split}
  \label{eq:l_relation}
\end{equation}
where $C$ is a constant. 
If the critical behavior of $l_s(t)$ is to be $l_s(t) = l_0 - A_\pm\abs{\delta t}^{\alpha_\pm} + o(\abs{\delta t}^{\alpha_\pm})$, then to satisfy Eq. \ref{eq:l_relation}, it has to be that $\alpha_+ = \alpha_- = \frac{1}{2}$.    
\section{An example of $\frac{1}{2}$-DQPT: the three-state Potts model}
\label{sec:Potts}
\begin{figure}[htb]
\centering
\begin{minipage}{.24\textwidth}
  \includegraphics[scale=0.27]{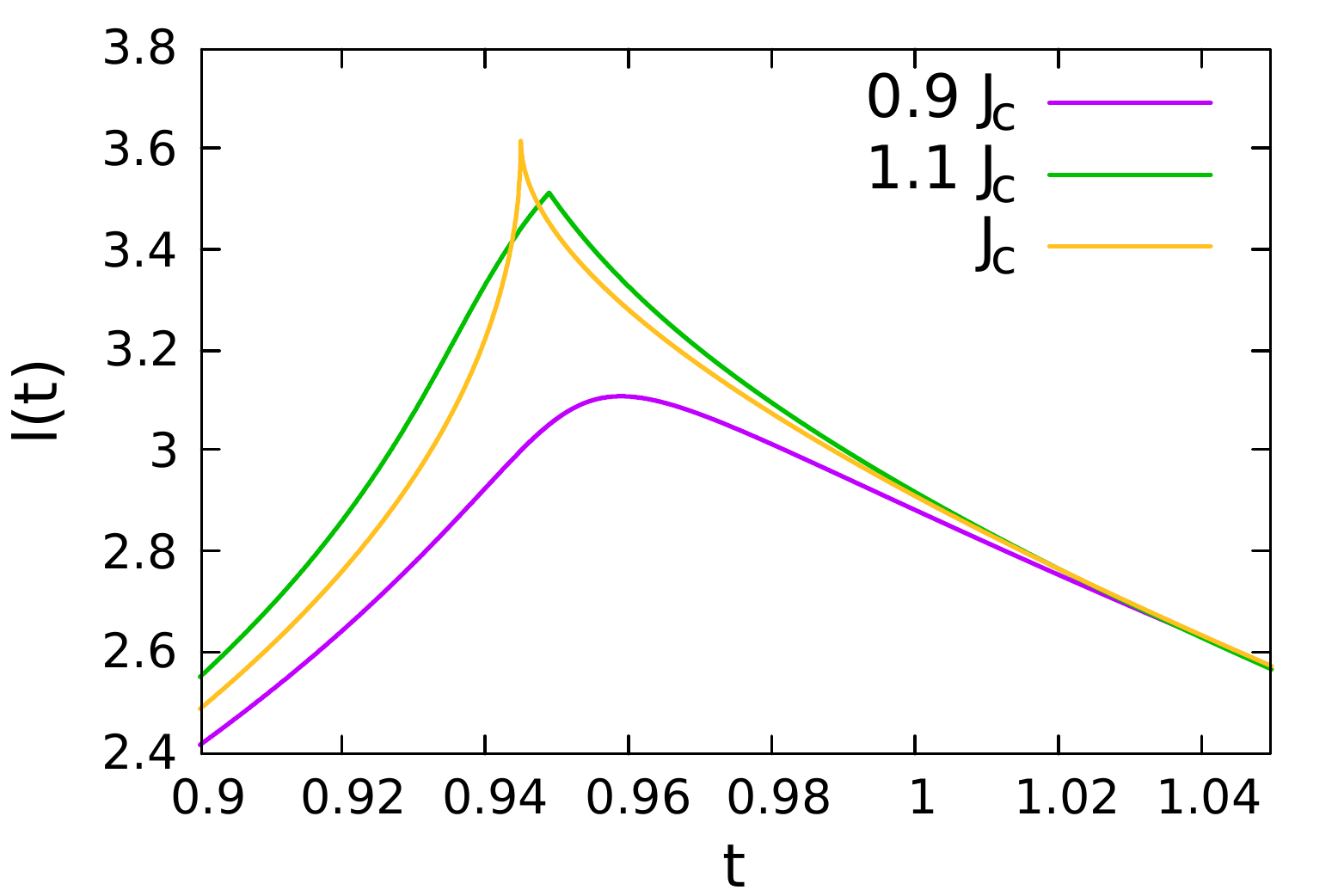}
\end{minipage}%
\begin{minipage}{.24\textwidth}
  \includegraphics[scale=0.27]{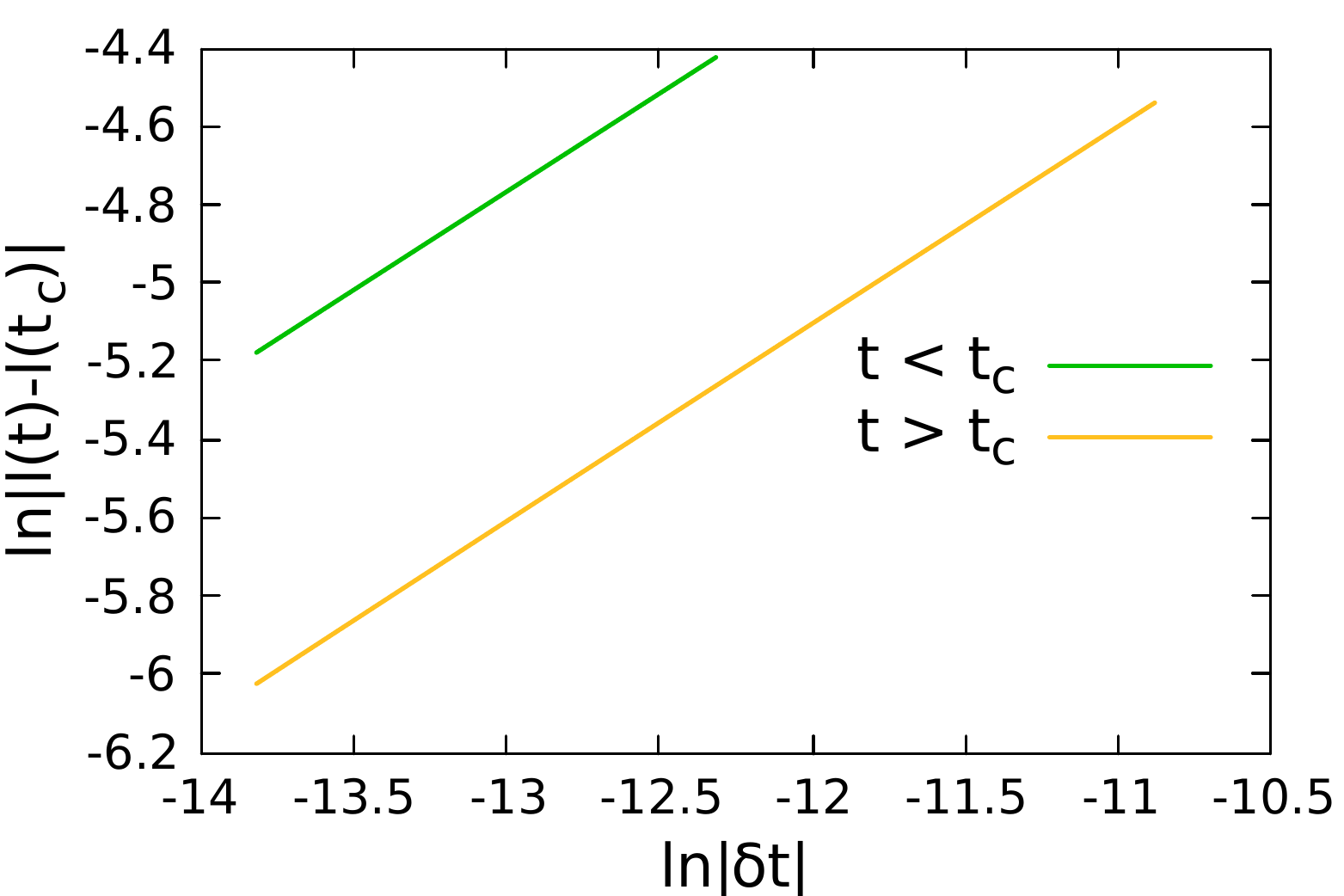}
\end{minipage}%
\caption{The rate function of the three-state Potts chain.
Left: The rate function for $J = 0.9 J_c, 1.1 J_c$ and $J_c$, obtained through the iTEBD algorithm. 
Right: The log-log plot of the rate function at $J_c$, for $t \in [t_c -4.832 \times 10^{-6}, t_c - 1\times 10^{-6}]$ and $t \in [t_c + 1\times 10^{-6}, t_c + 1.88 \times 10^{-5}]$, where $t_c = 0.9449044833$.
A linear-fit is performed on this plot to obtain the critical exponents. 
}
\label{fig:rate_itebd}
\end{figure}

Now that we have established that it is possible for $\frac{1}{2}$-DQPT to occur, does it actually happen? 
As mentioned, the place to look for such a DQPT is where linear-DQPTs disappear.  
Let us consider in detail an example in \cite{DQPT_Potts}, which studies the three-state Potts chain with the Hamiltonian,
\begin{equation}
  \hat H_{\Potts} = -J \sum_{i=1}^L (\hat \sigma_i^\dag \hat \sigma_{i+1} + \hat \sigma_{i+1}^\dag \hat \sigma_i) - \sum_{i=1}^L(\hat \tau_i^\dag + \hat \tau_{i}).
\end{equation}
The operators $\hat \sigma_i$ and $\hat \tau_i$ act on the three states of the local Hilbert space at site $i$, which we label by $\ket{0}_i, \ket{1}_i$, and $\ket{2}_i$.  
In this local basis, $\hat\sigma_i$ is a diagonal matrix with diagonal elements $\omega^s$ where $\omega = e^{i2\pi/3}$ and $s = 0, 1, 2$. 
$\hat\tau_i$ permutes $\ket{0}_i\rightarrow \ket{1}_i, \ket{1}_i \rightarrow \ket{2}_i$, etc., and together with $\hat \tau_i^\dag$ acts as a transverse-field.   
Off equilibrium criticality, this transfer matrix can be efficiently obtained with the time evolution algorithms \cite{itebd} based on MPS.     
In the Fig. 3 of \cite{DQPT_Potts}, its authors studied the Loschmidt rate function of the fully polarized ferromagnetic quenched state:
\begin{equation}
  \ket{\psi_0} = \otimes_i \ket{0}_i
\end{equation}
They found that as $J$ was varied from 0.03 to 0.1, the first peak of $l(t)$ changed from a smooth peak to a linear cusp.  
They also studied the rate function in the change-over region, but on a very limited parameter set, and concluded that the singularity seem still linear.   
Here we do a more refined scan of the parameter $J$ with the infinite-system time evolution block decimation algorithm (iTEBD) \cite{itebd} implemented in ITensor \cite{ITensor}.
A bond dimension of 10 is found to converge the calculation. 
The time step is set to be $10^{-4}$ before $t = 0.94485$ and $10^{-10}$ afterwards.  
The result of the calculation is given in Fig. \ref{fig:rate_itebd}. 
We find that the smooth peak changes to the linear cusp at approximately $J_c = 0.0572776316(1)$ with a critical time at $t_c = 0.9449044833(1)$. 

Here we pause to comment on what we mean by the convergence of the calculation.   
Of course, with a bond dimension of only 10, one cannot converge the calculation to the accuracy on the order of $10^{-10}$. 
Indeed, because the $J_c$ and $t_c$ are non-universal and depend on the details of the numerical approximations, one will find different values for them as one increases the bond dimension, presumably closer to the exact value defined by the quench protocol and the system Hamiltonian.   
However, the generality of the RG argument guarantees the universality and the robustness of the value of the critical exponent. 
This means that fixing a set of numerical approximations, i.e., the finite bond dimension and time step, one will always find a $t_c$ and $J_c$, for example, the $t_c$ and $J_c$ quoted above,  such that the linear-DQPT terminates, at which point Eq. \ref{eq:condition2} occurs and the DQPT has exponent $\frac{1}{2}$.  
If one increases the bond dimension, the $t_c$ and $J_c$ will take on different values, but the critical exponent of the DQPT will be the same. 
It is in this sense that we say we have converged our calculation. 

\begin{figure}[hth]
\centering
\begin{minipage}{.24\textwidth}
  \includegraphics[scale=0.27]{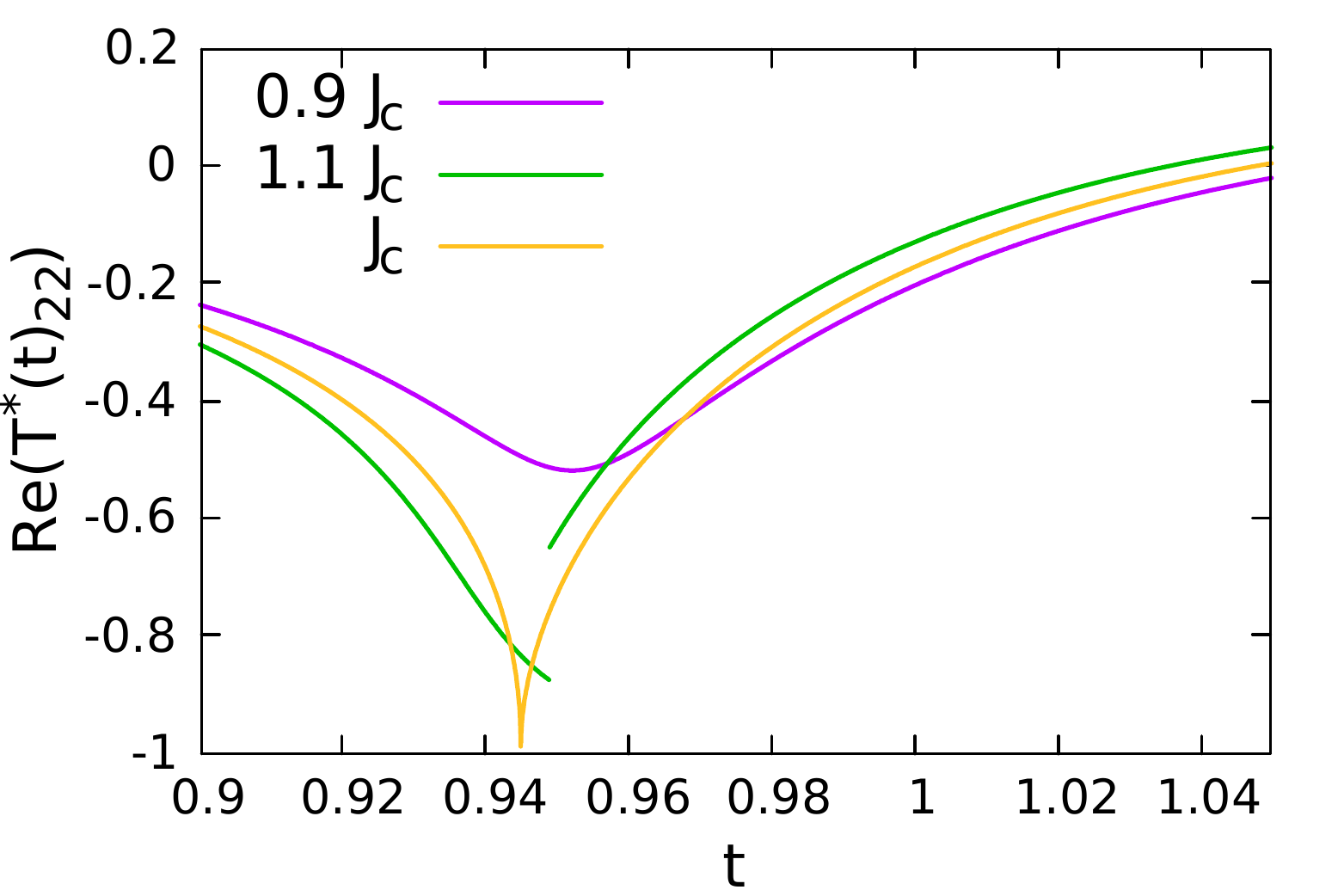}
\end{minipage}%
\begin{minipage}{.24\textwidth}
  \includegraphics[scale=0.27]{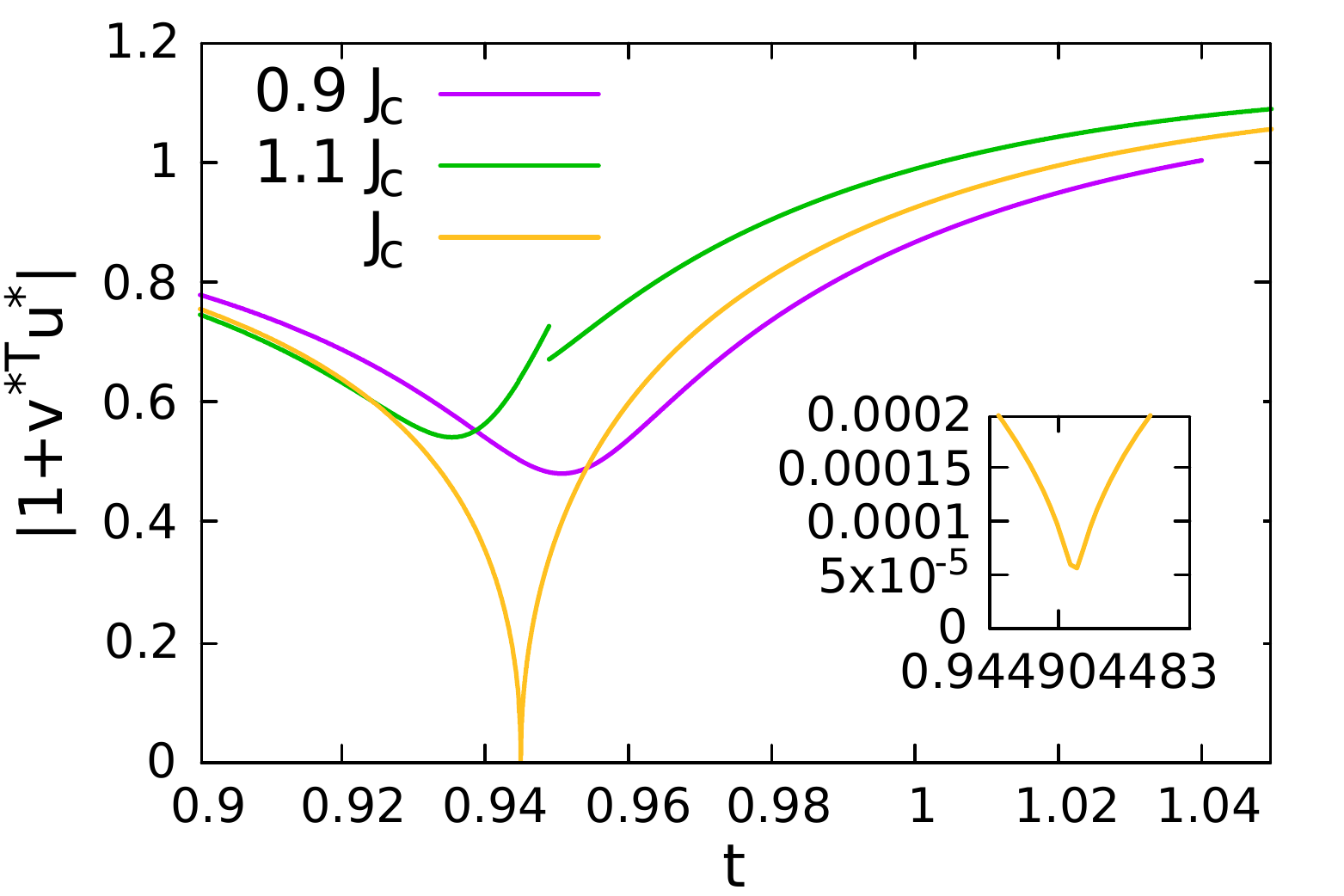}
\end{minipage}%
\caption{The fixed-point transfer matrix of the Potts chain for $J = 0.9 J_c, 1.1 J_c$ and $J_c$.
Left: The real part of the matrix element of $\T^*$ in row 2 and column 2. 
Right: The $\abs{1+\v^T\u}$ of $\T^*$.
The inset is a blowup at $J_c$ for $t$ near $t_c$.  
$\T^*$ is taken as the renormalized transfer matrix after $500$ RG iterations.   
}
\label{fig:rg_itebd}
\end{figure}

In Fig. \ref{fig:rate_itebd}, at $J_c$, the rate function at $t_c$ shows a singular cusp with a critical exponent which is numerically fitted to be $0.502$ and $0.505$ for $t < t_c$ and $t > t_c$ respectively (Fig. \ref{fig:rate_itebd}, right panel). 
Eq. \ref{eq:condition} is verified beyond the floating-point accuracy ($10^{-15}$) for $\T^\rg{500}(t)$ for all $t$ studied.  
Eq. \ref{eq:condition2} is also satisfied to very high precision: at the estimated critical parameter and time, $\abs{1+\v^{*T}\u^*}$ is found to be less than 0.0001 (Fig. \ref{fig:rg_itebd}).   
\section{Discussion}
\label{sec:Discussion}
In this paper, we have analyzed the RG procedure of the transfer matrix of the Loschmidt amplitude in detail. 
In particular, we have paid special attention to where the RG procedure itself becomes non-analytic. 
This gives a new RG fixed-point that controls a DQPT with exponent $\frac{1}{2}$.   
Such RG fixed-points occur in general where linear-DQPTs terminate. 
However, this does not mean that when linear-DQPTs disappear, there has to be a $\frac{1}{2}$-DQPT. 

\begin{figure}[htb]
\centering
\begin{minipage}{.16\textwidth}
  \includegraphics[scale=0.2]{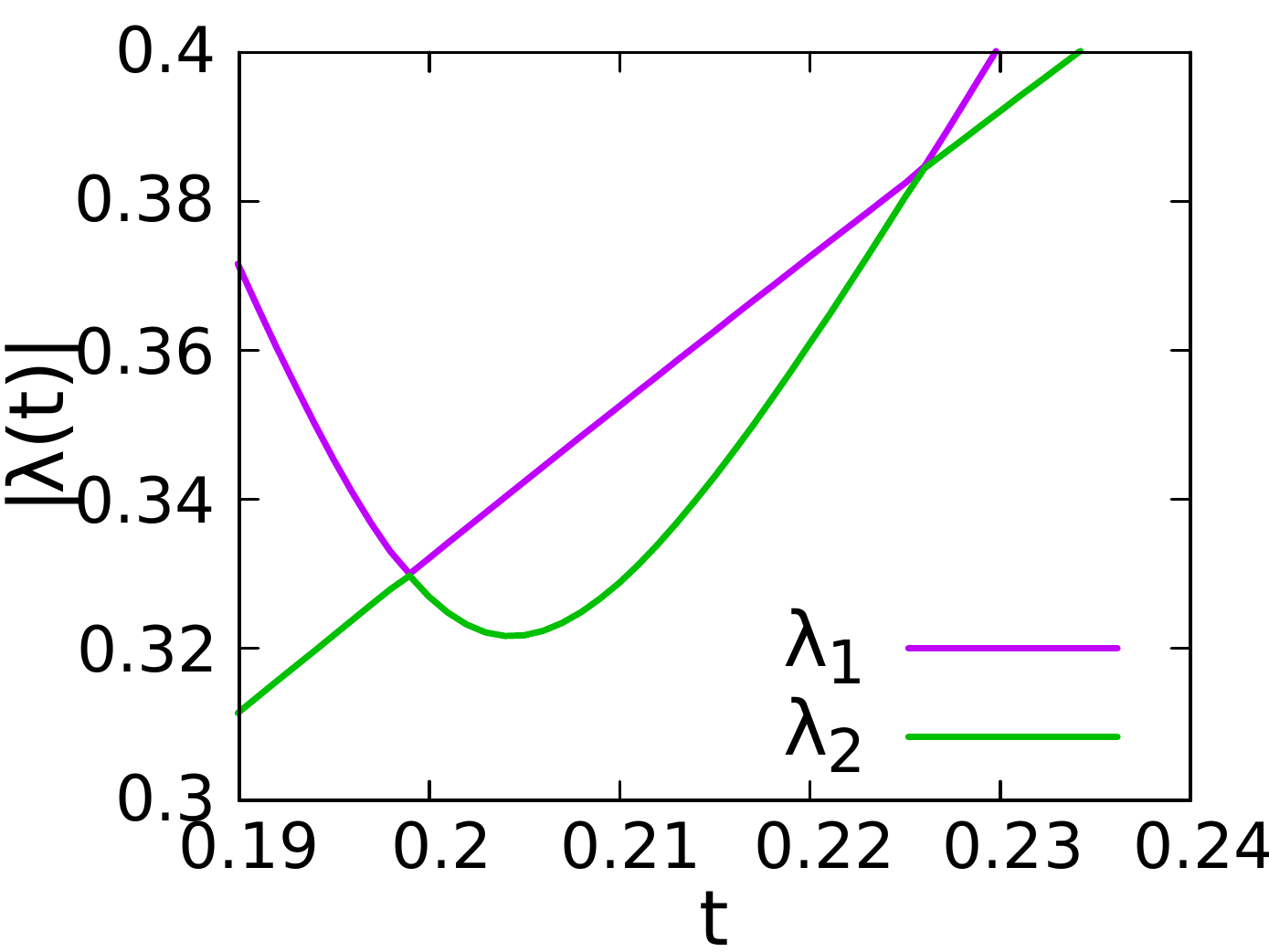}
\end{minipage}%
\begin{minipage}{.16\textwidth}
  \includegraphics[scale=0.2]{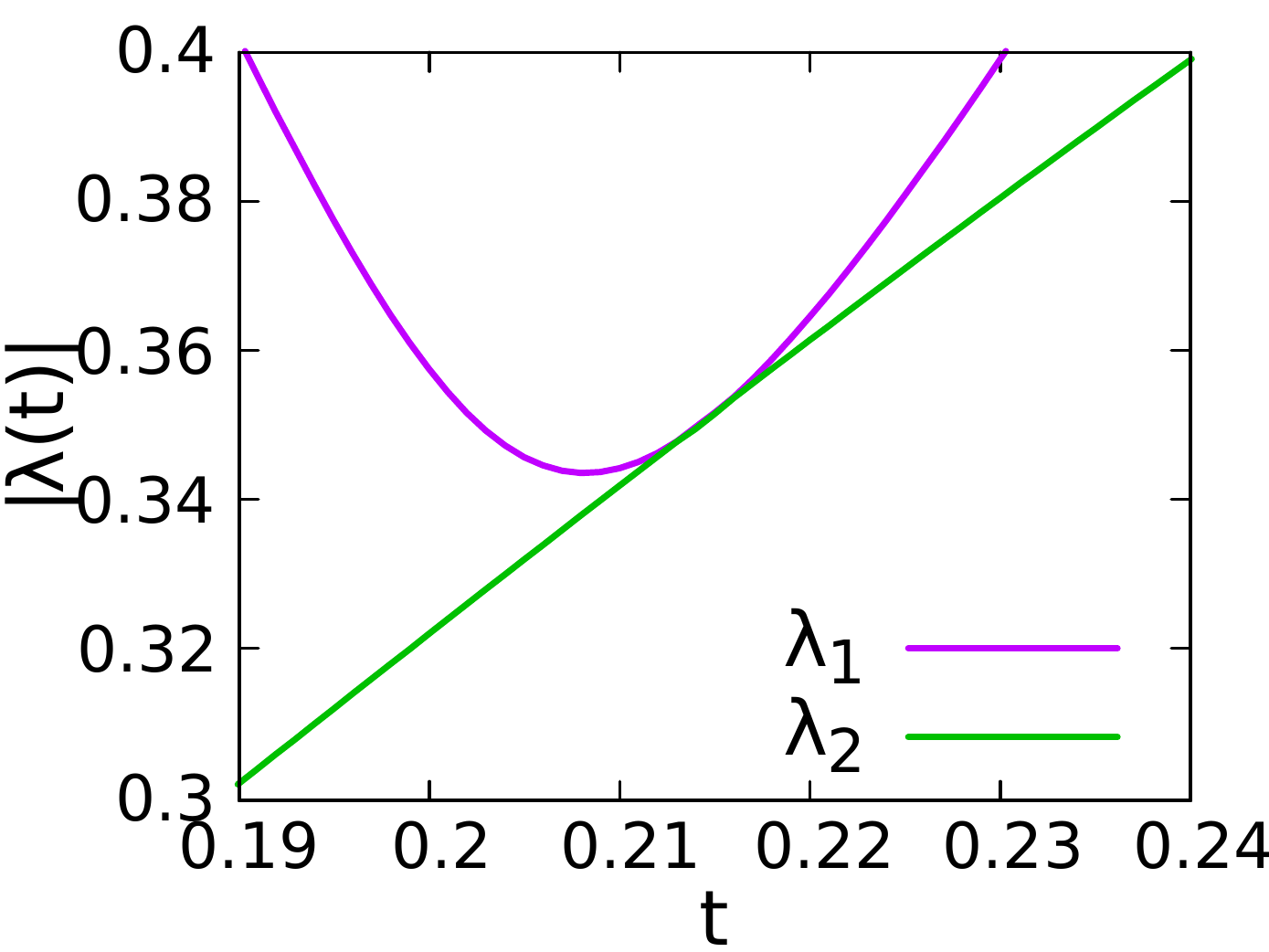}
\end{minipage}%
\begin{minipage}{.16\textwidth}
  \includegraphics[scale=0.2]{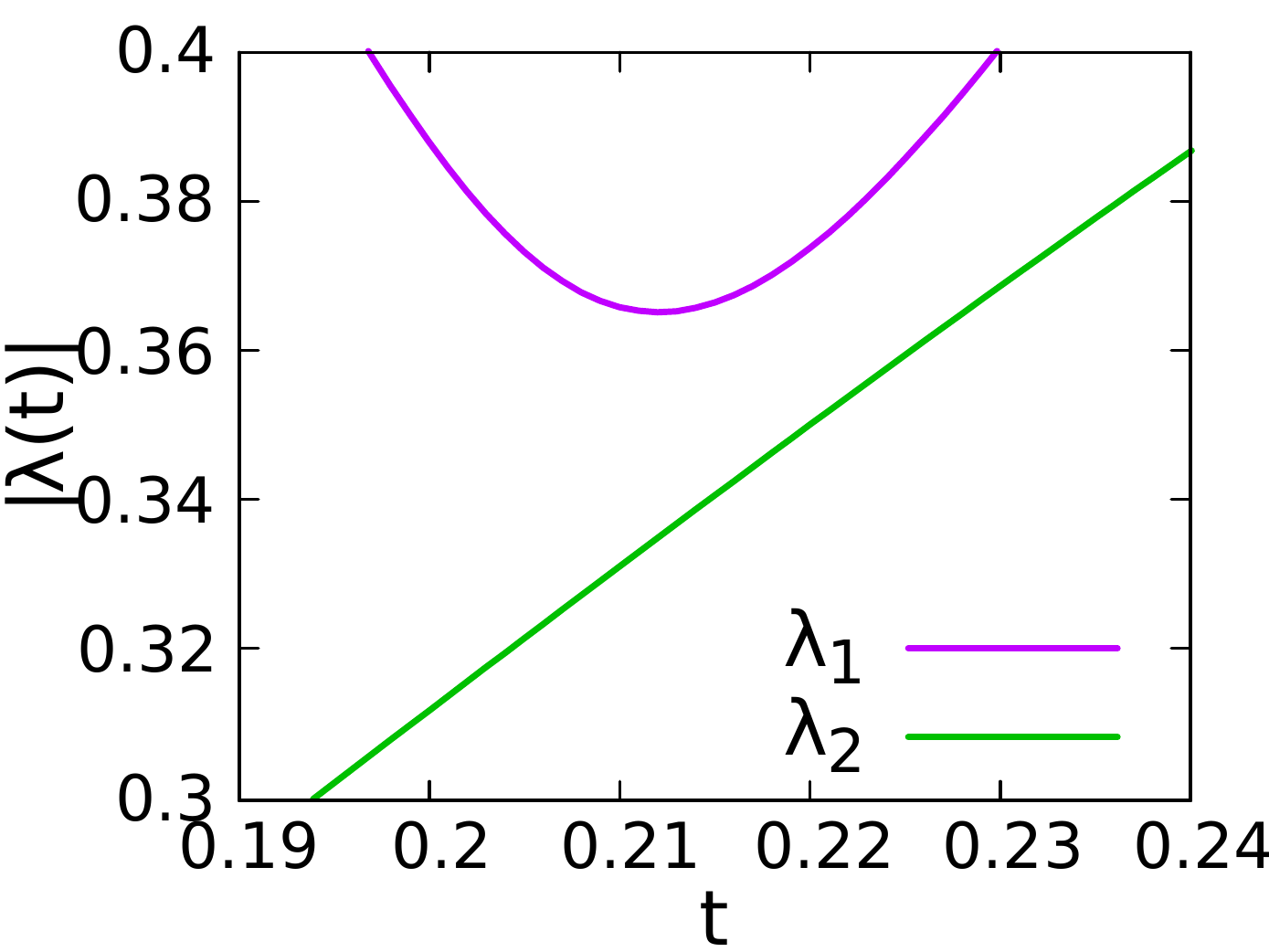}
\end{minipage}%
\caption{The modulus of the leading and the sub-leading eigenvalues, $\lambda_1$ and $\lambda_2$, of the transfer matrix of the XY Ising chain with $h_0 = 3, h_1 = 3, \gamma_0 = 3$. $\gamma_1 = \gamma_c - 0.1$ on the left, $\gamma_c$ in the middle, and $\gamma_c + 0.1$ on the right. 
The transfer matrix is obtained with iTEBD.
}
\label{fig:XY}
\end{figure}

For example, consider quenching the XY Ising chain with Hamiltonian \cite{DQPT_XY}, 
\begin{equation}
  \hat{H}_{\text{XY}} = \sum_{i=1}^N \frac{1+\gamma}{2} \hat\sigma_i^x\hat\sigma_{i+1}^x + \frac{1-\gamma}{2} \hat \sigma_i^y \hat \sigma_{i+1}^y - h \hat\sigma_i^z, 
\end{equation}
where $\sigma^{x,y,z}$ are the Pauli matrices. 
Let $\gamma_0$ and $h_0$ be the parameters of the pre-quenched Hamiltonian, and $\gamma_1$ and $h_1$ that of the post-quenched Hamiltonian.  
Then when $h_0 = 3, h_1 = 3, \gamma_0 = 3$, and $\gamma_1 < \gamma_c = -8/3$, linear-DQPTs occur.
When $\gamma_1 > \gamma_c$, linear-DQPTs disappear \cite{DQPT_XY}. 
The leading and sub-leading eigenvalues of the transfer matrix are plotted in Fig. \ref{fig:XY}. 
There, clearly, at the termination point of the linear-DQPT, $l(t)$ will be a smooth function of time.  
In addition, the linear-DQPT in Fig. \ref{fig:XY} is characterized by double cusps before the termination, whereas if a linear-DQPT disappears into a $\frac{1}{2}$-DQPT, it disappears alone. 

There thus seems to be a qualitative difference in how linear-DQPTs terminate between the Potts example and the XY Ising example.   
While the termination of linear-DQPTs shown in Fig. \ref{fig:XY} is in a sense accidental, the $\frac{1}{2}$-DQPT seems much more non-trivial. 
It would be very nice if one can further clarify this difference in future works. 
\section{Appendix}
\subsection{Proof of Eq. \ref{eq:ls_w} and Eq. \ref{eq:w_recursion}}
\label{sec:proof_ls_w}
We start from Eq. \ref{eq:ls}:
\begin{equation}
  \begin{split}
  l_s(t) &= \sum_{n=n_0}^\infty \frac{-1}{2^n} \ln\abs{\v_c^T\delta\u^\rg{n}(t) + (\delta\v^\rg{n}(t))^T\u_c}.
\end{split}
\end{equation}
This prompts us to simplify Eq. \ref{eq:recursion} by defining $\delta w = \frac{1}{2}(\v_c^T \delta\u + \delta\v^T \u_c)$ and $\delta x = \v_c^T\delta\X\u_c$, which have the following recursion relation  
\begin{equation}
\begin{split}
  \delta w^\rg{n+1} &= \delta w^\rg{n} + \frac{\delta x^\rg{n}}{2\delta w^\rg{n}}
  \\
  \delta x^\rg{n+1} &= -\left(\frac{\delta x^\rg{n}}{2\delta w^\rg{n}}\right)^2
\end{split}
\label{eq:recursion2}
\end{equation}
Quite remarkably, Eq. \ref{eq:recursion2} has a conservative quantity:  
\begin{equation}
  \Delta x \equiv \delta x^\rg{n+1} + (\delta w^\rg{n+1})^2 = \delta x^\rg{n} + (\delta w^\rg{n})^2
\end{equation}
Because $\Delta x$ depends only on $\T$ at the starting point of the RG flow, it is an analytic function of $t$.    
At criticality, in addition, because $\delta x^\rg{n}$ and $\delta w^\rg{n}$ both tend to zero as $n \rightarrow \infty$, $\Delta x$ must be zero to start with.    
One can thus write to the leading order of $\delta t\equiv  t - t_c$, 
\begin{equation}
  \Delta x(t) = a \delta t
\end{equation}
where $a$ is a non-universal constant. 

Replacing $\delta x^\rg{n}$ by $\Delta x$, one obtains: 
\begin{equation}
  \delta w^\rg{n+1} = \frac{1}{2}(\delta w^\rg{n} + \frac{a\delta t}{\delta w^\rg{n}}), \hspace{2mm} \delta w^\rg{n_0}(t) = \delta w_0 
\end{equation}
where we have noted that $\delta w^\rg{n_0}$ is an analytic function of $t$ and is a constant $\delta w_0$ to the leading order of $\delta t$. 
Making one last re-definition $\delta \tilde{w} \equiv \frac{\delta w}{\sqrt{a\delta t}}$ and rewriting Eq. \ref{eq:ls}, one finally obtains a set of equations simple enough to extract the critical behavior of $l_s(t)$:   
\begin{equation}
  \begin{split}
    \delta \tilde w^\rg{n+1} &= \frac{1}{2}(\delta \tilde w^\rg{n} + \frac{1}{\delta \tilde w^\rg{n}}), \hspace{2mm} \delta \tilde w^\rg{n_0}(t) = \frac{\delta w_0}{\sqrt{a\delta t}} 
    \\
    l_s(t) &= \sum_{n=n_0}^\infty \frac{-1}{2^{n}}\ln\abs{2\sqrt{a\delta t}\delta\tilde w^\rg{n}(t)}
    \\
    &\sim \frac{-1}{2^{n_0-1}}\ln\abs{\sqrt{\delta t}} + \sum_{n=n_0}^\infty \frac{-1}{2^{n}} \ln \abs{\delta \tilde\omega^\rg{n}(t)}
\end{split}
\end{equation}
where we have again dropped analytic parts from $l_s(t)$. 
This completes the proof for Eq. \ref{eq:ls_w} and Eq. \ref{eq:w_recursion}. 

\begin{acknowledgments}
The author is grateful to Ling Wang for hosting him at the Beijing Computational Science Research Center, introducing him to DQPTs, and many stimulating discussions.  
He is also grateful for mentorship from his advisor Roberto Car at Princeton. 
The author acknowledges support from the DOE Award DE-SC0017865. 

\end{acknowledgments}
\bibliography{abc}

\begin{thebibliography}{44}
\bibitem [{def()}]{defective}%
  \BibitemOpen
  \href@noop {} {}\bibinfo {note} {If a matrix is diagonalizable, then for any
  eigenvalue, the right eigenvector and the left eigenvector cannot be
  orthogonal.}\BibitemShut {Stop}%
\end{thebibliography}
\end{document}